\documentclass[aps,prd,twocolumn,10pt,superscriptaddress,nofootinbib,longbibliography]{revtex4-1}

\usepackage{amssymb}
\usepackage{graphicx}
\usepackage{amsmath}
\usepackage{hyperref}
\usepackage{multirow}
\usepackage{setspace}
\usepackage{verbatim}
\usepackage{float}
\usepackage{color}
\usepackage{ulem}
\usepackage[utf8]{inputenc}
\usepackage[table,xcdraw]{xcolor}
\usepackage{makecell}
\usepackage{url}
\usepackage{bm}

\begin{document}

\title{Resolving the Planck-DESI tension by nonminimally coupled quintessence}

\author{Jia-Qi Wang}
\email{wangjiaqi@itp.ac.cn}
\affiliation{Institute of Theoretical Physics, Chinese Academy of Sciences (CAS), Beijing 100190, China}
\affiliation{University of Chinese Academy of Sciences (UCAS), Beijing 100049, China}

\author{Rong-Gen Cai}
\email{caironggen@nbu.edu.cn}
\affiliation{Institute of Fundamental Physics and Quantum Technology, Ningbo University, Ningbo, 315211, China}

\author{Zong-Kuan Guo}
\email{guozk@itp.ac.cn}
\affiliation{Institute of Theoretical Physics, Chinese Academy of Sciences (CAS), Beijing 100190, China}
\affiliation{University of Chinese Academy of Sciences (UCAS), Beijing 100049, China}
\affiliation{School of Fundamental Physics and Mathematical Sciences, Hangzhou Institute for Advanced Study, University of Chinese Academy of Sciences, Hangzhou 310024, China}

\author{Shao-Jiang Wang}
\email{schwang@itp.ac.cn (Corresponding author)}
\affiliation{Institute of Theoretical Physics, Chinese Academy of Sciences (CAS), Beijing 100190, China}
\affiliation{Asia Pacific Center for Theoretical Physics (APCTP), Pohang 37673, Korea}

\begin{abstract}
The Planck measurement of the cosmic microwave background (CMB) has established the $\Lambda$-cold-dark-matter ($\Lambda$CDM) model as the concordant model along with other observations. However, recent measurements of baryon acoustic oscillations (BAO) from the Dark Energy Spectroscopic Instrument (DESI) have renewed the matter fraction $\Omega_\mathrm{m}$ tension between Planck-$\Lambda$CDM and DESI-$\Lambda$CDM. Directly reconciling this CMB-BAO tension with a dynamical DE in Chevallier-Polarski-Linder (CPL) parametrization seems to imply a crossing of the equation-of-state (EOS) through $w=-1$ at low redshifts. In this paper, we resolve this $\Omega_\mathrm{m}$ tension by allowing for the DM nonminimally coupled to gravity via a quintessence field. This non-minimal coupling is preferred over $3\sigma$ confidence level. Consequently, even though the usual effective EOS of the coupled quintessence apart from the standard CDM part never crosses but always is above $w=-1$, a misidentification with the $w_0w_a$CDM model would exactly fake such a crossing behavior, and the tensions on neutrino mass and growth rate in the $\Lambda$CDM model are also relieved in our model as a result of the resolved $\Omega_\mathrm{m}$ tension.
\end{abstract}

\maketitle

\section{Introduction}\label{sec:intro}
The Planck measurement of cosmic microwave background (CMB)~\cite{Planck:2018vyg}, along with the completed Sloan Digital Sky Survey of baryon acoustic oscillations (BAO)~\cite{eBOSS:2020yzd} and the PantheonPlus compilation of Type Ia supernovae (SNe Ia)~\cite{Brout:2022vxf}, all agree roughly on the same parameter region of the $\Lambda$-cold-dark-matter ($\Lambda$CDM) model. However, the recent data release 2 (DR2) of BAO results from three-year (Y3) observations with Dark Energy Spectroscopic Instrument (DESI)~\cite{DESI:2025zgx}, when combined with both Planck-CMB and the five-year compilation of the Dark Energy Survey (DESY5)~\cite{DES:2024jxu} of SNe Ia, has claimed more than $4\sigma$ deviation~\cite{DESI:2025zgx} from $\Lambda$CDM within Chevallier-Polarski-Linder (CPL) parametrization $w=w_0+w_a(1-a)$~\cite{Chevallier:2000qy,Linder:2002et} on the equation of state (EOS) of dynamical dark energy (DDE)~\cite{Peebles:2002gy}. Although the inclusion of DESY5 compilation, especially its low-$z$ sample, has been questioned~\cite{Efstathiou:2024xcq,Huang:2025som,Zhong:2025gyn} for their distinct behaviors from the PantheonPlus compilation, the Planck+DESI combination alone without low-$z$ sample or even without the whole DESY5 compilation still prefers a DDE with a significance exceeding $2\sim3\sigma$~\cite{DESI:2025zgx}. 

However, when constraining the matter fraction today $\Omega_\mathrm{m}$ in the $\Lambda$CDM model, there is a mild discrepancy ($1.8\sigma$) between Planck-CMB ($\Omega_\mathrm{m}=0.3169\pm0.0065$) and DESI-BAO ($\Omega_\mathrm{m}=0.2975\pm0.0086$) constraints~\cite{DESI:2025zgx}. Moreover, this $\Omega_\mathrm{m}$ discrepancy even becomes a considerable tension ($2.3\sigma\sim3.6\sigma$) in the $w_0w_a$CDM model also between Planck-CMB ($\Omega_\mathrm{m}=0.220_{-0.078}^{+0.019}$) and DESI-BAO ($\Omega_\mathrm{m}=0.352_{-0.018}^{+0.041}$) constraints~\cite{DESI:2025zgx}. A similar $\Omega_\mathrm{m}$ tension is still persistent ($2.9\sigma$ and $2.4\sigma\sim3\sigma$) between DESI BAO and the DESY5 constraints~\cite{DES:2024jxu} for both the $\Lambda$CDM ($\Omega_\mathrm{m}=0.352\pm0.017$) and $w_0w_a$CDM ($\Omega_\mathrm{m}=0.495_{-0.043}^{+0.033}$) models, respectively. 
Therefore, this $\Omega_\mathrm{m}$ tension is more plausibly alleviated by the reduced constraining power in the $w_0w_a$CDM model, rather than being completely resolved.
Nevertheless, even though the DESI BAO alone still prefers a DDE but only at  $1.7\sigma$~\cite{DESI:2025zgx}, the crossing point can still be constrained around the redshift $z=0.45_{-0.05}^{+0.03}$~\cite{Ye:2025ark} from the degeneracy direction of $w_0$ and $w_a$. This crossing behavior seems to be also robust to nonparametric reconstructions~\cite{Jiang:2024xnu,DESI:2025wyn} and non-DESI data constraints~\cite{Park:2024vrw}.

Therefore, any satisfactory resolution to this $\Omega_\mathrm{m}$ tension~\cite{Colgain:2024xqj,Colgain:2024ksa,Colgain:2024mtg,Wang:2025bkk,Chaudhary:2025pcc,Lee:2025kbn} should also reproduce the crossing behavior as well, but a simple $w_0w_a$CDM model does not meet this criterion. Since a single perfect fluid minimally coupled to Einstein gravity cannot realize a smooth crossing behavior~\cite{Vikman:2004dc,Deffayet:2010qz}, a recent trend in explaining the DESI results tends to modify the Einstein gravity~\cite{Ye:2024ywg,Pan:2025psn,Cai:2025mas}, especially a nonminimally coupled (dark) matter sector to Einstein gravity via a quintessence field~\cite{Gomez-Valent:2020mqn,Cai:2021wgv,Yu:2022wvg,Karwal:2021vpk,Pitrou:2023swx,Uzan:2023dsk,Wolf:2024stt,Ye:2024zpk,Tiwari:2024gzo,Chakraborty:2025syu,Khoury:2025txd,Wolf:2025jed,Bedroya:2025fwh,Brax:2025ahm}. Similar dark matter-dark energy (DM-DE) interactions~\cite{Chakraborty:2024xas,Wang:2024hks,Giare:2024smz,Li:2024qso,Aboubrahim:2024cyk,Li:2025owk,Sabogal:2025mkp,Tsedrik:2025cwc,Zhai:2025hfi,Shah:2025ayl,Silva:2025hxw,Pan:2025qwy,Yashiki:2025loj,Barman:2025ryg,Li:2025ula} have recently been shown to reproduce the DESI-preferred crossing behavior.

In this paper, we propose to solve the aforementioned $\Omega_\mathrm{m}$ tension using a nonminimally coupled quintessence (NMCQ) model~\cite{Amendola:1999er,Wetterich:1994bg,Khoury:2003aq,Khoury:2003rn,Upadhye:2012vh} with, in specific, the Ratra-Peebles potential~\cite{Ratra:1987rm,Peebles:1987ek} and dilaton coupling~\cite{Wetterich:1987fm,Damour:1994zq}, commonly arising from dimensional reductions of string theories and consistent with swampland criteria~\cite{Svrcek:2006yi,Ooguri:2016pdq,Agrawal:2018own,Ooguri:2018wrx,Bedroya:2025fwh}. 
Such a coupling induces an evolving DM mass and dynamically shifts the background evolution, thereby reconciling the lower $\Omega_\mathrm{m}$ inferred from DESI BAO with the higher value from Planck CMB. The crossing behavior is not a real physical effect but emerges as a mismatched modeling with the $w_0w_a$CDM model; even the effective EOS of our coupled quintessence (after excluding the standard CDM part) never crosses $w=-1$. This is different from other interacting DE-DM models with their effective EOS indeed crossing $w=-1$.

\section{The NMCQ model}\label{sec:model}
The action of the NMCQ model is described by the action $S=S_\mathrm{GR}+S_\mathrm{SM}+S_\mathrm{DM}+S_\varphi$, where $S_\mathrm{GR}=\int\mathrm{d}^4x\sqrt{-g}\,M_\mathrm{Pl}^2R/2$ is the usual Einstein-Hilbert action, while the standard-model (SM) particles $\psi_\mathrm{SM}$ are minimally coupled to Einstein gravity by $S_\mathrm{SM}=\int\mathrm{d}^4x\mathcal{L}_\mathrm{SM}[\psi_\mathrm{SM};g_{\mu\nu}]$, but the DM sector $\psi_\mathrm{DM}$ is nonminimally coupled to Einstein gravity by $S_\mathrm{DM}=\int\mathrm{d}^4x\mathcal{L}_\mathrm{DM}[\psi_\mathrm{DM};\tilde{g}_{\mu\nu}\equiv\mathcal{A}^2(\varphi)g_{\mu\nu}]$ via a scalar field,
\begin{align}
  S_\varphi &= \int \mathrm{d}^4x \sqrt{-g} \left[-\frac{1}{2} g^{\mu\nu}(\varphi) \partial_\mu \varphi \partial_\nu \varphi - V(\varphi) \right].
\end{align}
A simple but representative configuration is to consider a dilaton coupling $\mathcal{A}(\varphi)=\mathrm{e}^{-\beta\varphi / M_\mathrm{Pl}}$~\cite{Wetterich:1987fm,Damour:1994zq,Bedroya:2025fwh} and the Ratra-Peebles potential $V(\varphi)=\alpha \Lambda^4(\varphi/M_\mathrm{Pl})^{-n}$~\cite{Ratra:1987rm,Peebles:1987ek}. The $\Lambda$CDM model is recovered at $\beta=n=0$. This model is not aimed at solving the cosmological constant problem~\cite{Weinberg:1988cp}, and hence, we will simply set $\Lambda^4\equiv3M_\mathrm{Pl}^2H_0^2$ at the current critical energy density for an $\mathcal{O}(1)$ coefficient $\alpha$. Here, $H_0\equiv100h$ km/s/Mpc is the Hubble constant, and $M_\mathrm{Pl}\equiv1/\sqrt{8\pi G}$ is the reduced Planck mass. The scalar-mediated fifth force only acts on the DM component, thus remaining undetected by current experiments.

Varying the total action with respect to the Einstein-frame Friedmann-Lema\^{i}tre-Robertson-Walker (FLRW) metric $g_{\mu\nu}$, scalar $\varphi$, and DM $\psi_\mathrm{DM}$ leads to the following equations of motions (EOMs)~\cite{Amendola:1999er,Wetterich:1994bg,Khoury:2003aq,Khoury:2003rn,Upadhye:2012vh}:
\begin{align}
\rho_\mathrm{r}+\rho_\mathrm{b}+\rho_\mathrm{DM}+\rho_\varphi&=3M_\mathrm{Pl}^2H^2,\\
\dot{\rho}_\varphi+3H(1+w_\varphi)\rho_\varphi&=-\frac{\mathcal{A}'(\varphi)}{\mathcal{A}(\varphi)}\dot{\varphi}\rho_\mathrm{DM}\label{Eq.EOMphi},\\
\dot{\rho}_\mathrm{DM}+3H\rho_\mathrm{DM}&=+\frac{\mathcal{A}'(\varphi)}{\mathcal{A}(\varphi)}\dot{\varphi}\rho_\mathrm{DM}\label{Eq.EOMDM},
\end{align}
where the evolution of SM fields with the scale factor $a$ (after setting $a_0\equiv1$) is standard for both radiations $\rho_\mathrm{r}=\rho_{\mathrm{r},0}a^{-4}$ and baryons $\rho_\mathrm{b}=\rho_{\mathrm{b},0}a^{-3}$, and the scalar-field EOS is defined as usual $w_\varphi\equiv p_\varphi/\rho_\varphi$ from the scalar pressure $p_\varphi=\frac12\dot{\varphi}^2-V(\varphi)$ and scalar density $\rho_\varphi=\frac12\dot{\varphi}^2+V(\varphi)$. The above DM-$\varphi$ coupling term does not render a standard $a^{-3}$ evolution for both Einstein-frame DM sector $\rho_\mathrm{DM}$ and Jordan-frame DM sector $\tilde{\rho}_\mathrm{DM}\equiv\mathcal{A}^{-4}(\varphi)\rho_\mathrm{DM}$. It turns out that it is this combination $\mathcal{A}^{-1}(\varphi)\rho_\mathrm{DM}\equiv\rho_\mathrm{CDM}=\rho_{\mathrm{CDM},0}a^{-3}$ that evolves as the standard CDM. We therefore define $\rho_{\mathrm{DM},0}\equiv\mathcal{A}(\varphi_0)\rho_{\mathrm{CDM},0}$ to yield
\begin{equation}\label{Eq.rhoDM}
  \frac{\rho_\mathrm{DM}}{\rho_{\mathrm{DM},0}}=\left(\frac{a}{a_0}\right)^{-3}\times \left(\frac{\mathcal{A}}{\mathcal{A}_0}\right).
\end{equation}

When solving EOMs, subtleties arise for the choices of initial condition and matching condition at the present day, as shown in the appendixes. The initial condition is secured by an attractor solution of the scalar field converging at $z=10^9$, and the matching condition  $\rho_{\mathrm{r},0}+\rho_{\mathrm{b},0}+\rho_{\mathrm{DM},0}+\frac12\dot{\varphi}_0^2+V(\varphi_0)=3M_\mathrm{Pl}^2H_0^2\equiv\rho_{\mathrm{crit},0}$ is realized by simultaneously shooting for both $\varphi_0$ and $\alpha$ values in terms of other observables $\Omega_i\equiv\rho_{i,0}/\rho_{\mathrm{crit},0}$ for $i=\mathrm{r}, \mathrm{b}, \mathrm{c}(\equiv\mathrm{CDM}), \mathrm{DM},\mathrm{m}(\equiv\mathrm{b+DM})$.

\section{Methodology and data}\label{sec:method}
We implement the data analysis for the $\mathrm{\Lambda}$CDM, $w_0w_a$CDM, and NMCQ models with a modified version of the cosmological linear Boltzmann code \texttt{CAMB}~\cite{Lewis:1999bs, Li:2023fdk, Li:2014eha, Hu:2013twa} to adapt to the nonminimal coupling case~\cite{Li:2023fdk, Li:2014eha}, and use the publicly available sampling code~\texttt{Cobaya}~\cite{Torrado:2020dgo,2019ascl.soft10019T} to perform Markov chain Monte Carlo (MCMC) analyses. The datasets include:
\begin{enumerate}
    \item[i.] \textbf{Planck 2018 CMB}: (i) the \textsc{CamSpec} version of \textsc{Planck} PR4 \textsc{NPIPE} high-multipole ($\ell > 30$) angular power spectra of temperature and polarization (TTTEEE) anisotropies~\cite{Rosenberg:2022sdy}; (ii) the low-multipole ($2 \leq \ell \leq 30$) temperature ($C_\ell^{TT}$) spectra extracted by \textsc{Commander}~\cite{Planck:2019nip}; (iii) the low-multipole E-mode polarization ($2 \leq \ell \leq 30$) power spectrum $C_\ell^{EE}$ using \textsc{SimAll} likelihood~\cite{Planck:2019nip}; (iv) CMB lensing data using \textsc{NPIPE} PR4 Planck reconstruction~\cite{Carron:2022eyg}.
    \item[ii.] \textbf{DESIY3 DR2 BAO}: The recent DESI Y3 BAO measurements of galaxies, quasars, and Lyman-$\alpha$ forest in Table IV of the DR2 paper~\cite{DESI:2025zgx}. 
    \item[iii.] \textbf{DESY5 SNe Ia}: the DESY5 compilation including 194 external low-redshift ($z\leq 0.1$) samples and 1635 high-redshift ($0.1<z<1.3$) DES-SN samples~\cite{DES:2024jxu}.
    \item[iv.] \textbf{$f\sigma_8$}: measurements of the product of the growth rate and the amplitude of linear matter fluctuations on a comving scale of $8h^{-1}$ Mpc, $f(z)\sigma_8(z)$, from peculiar velocity and redshift-space distortion (RSD) data~\cite{Said:2020epb,Beutler:2012px,Huterer:2016uyq,Boruah:2019icj,Turner:2022mla,Blake:2011rj,Blake:2013nif,Howlett:2014opa,Okumura:2015lvp,Pezzotta:2016gbo,eBOSS:2020yzd}. This dataset will be only used for a $\chi^2$ test but play no role in determining the cosmological parameters.
\end{enumerate}
We also used the PantheonPlus sample~\cite{Brout:2022vxf} and extended Baryon Oscillation Spectroscopic Survey (eBOSS) DR16 BAO~\cite{eBOSS:2020yzd} for comparison in the appendixes.

We sample two model parameters $\left\{n, \beta\right\}$ or $\left\{w_0, w_a\right\}$ and two cosmological parameters $\left\{\Omega_c, H_0\right\}$ when only the DESI or DES dataset was used. If CMB likelihoods were included, all the external priors would be flat. The details of the sampling and methods are similar to previous works, and will be stated in the appendixes.



\begin{table}[t]
  \centering
  \caption{Cosmological constraints on model parameters (``Para.'') in the $\Lambda$CDM, $w_0w_a$CDM, and NMCQ models from Planck+DESI+DESY5. The last two lines present the relative $\chi^2$-test and relative Bayes factor, $\ln \mathcal{B}_{ij}= \ln Z_i-\ln Z_\mathrm{\Lambda CDM}$ for the $w_0w_a$CDM and NMCQ models with respect to the $\Lambda$CDM model.}
  \renewcommand{\arraystretch}{1.25}
  \label{tab:cosmo_params}
  {\footnotesize
  \begin{tabular}{llll}
  \hline\hline
  Para. & $\Lambda$CDM & $w_0w_a$CDM & NMCQ \\
  \hline
  $\Omega_{\mathrm{b}}h^2$ & $0.02229\pm0.00011$ & $0.02224\pm0.00012$ & $0.02216\pm0.00012$ \\
  $\Omega_{\mathrm{c}}h^2$ & $0.1180\pm0.0006$ & $0.1191\pm0.0008$ & $0.1141\pm0.0013$ \\
  $100h$ & $67.99\pm0.27 $  & $66.74^{+0.55}_{-0.56}$ & $67.29\pm0.55$ \\
  $w_0$ & ... & $-0.756\pm0.057$ & ... \\
  $w_a$ & ... & $-0.840^{+0.220}_{-0.225}$ & ... \\
  $n$ & ... & ... & $0.62\pm0.18$ \\
  $\beta$ & ... & ... & $0.054^{+0.012}_{-0.008}$ \\
  \hline
  $\Omega_{\mathrm{m}}$ & $0.305\pm0.0034$ & $0.319\pm0.0055$ & $0.302\pm0.0053$ \\
  $S_8$ & $0.813\pm0.007$  & $0.827\pm0.009$ & $0.820\pm0.008$ \\
  \hline
  $\Delta \chi^2$ & $0$ & $-17.9$ & $-12.4$ \\
  $\ln \mathcal{B}_{ij}$ & $0$ & $+3.69\pm0.30$ & $+2.66\pm0.30$ \\
  \hline\hline
  \end{tabular}
  }
\end{table}

\begin{figure}[h]
  \includegraphics[width=\linewidth]{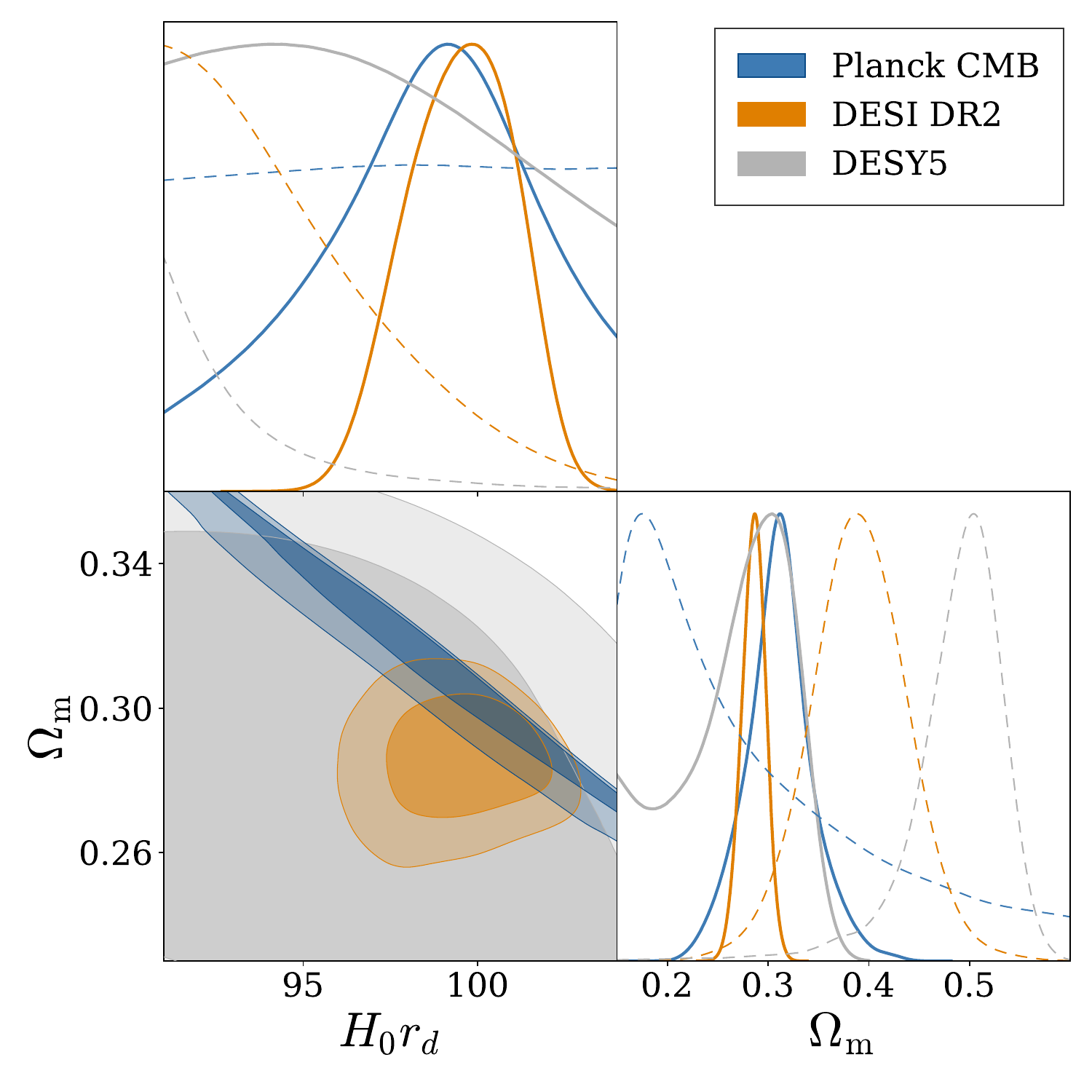}
  \caption{Cosmological constraints on $\Omega_\mathrm{m}$ and $H_0r_d$ in $w_0w_a$CDM (dotted) and NMCQ (solid) models from Planck CMB, DESI BAO, and DESY5 SNe, separately.}\label{fig:omegam} 
\end{figure}

\section{Cosmological constraints.}\label{sec:results}
The combined constraints from Planck+DESI+DESY5 for the $\Lambda$CDM, $w_0w_a$CDM, and NMCQ models are presented in Table~\ref{tab:cosmo_params} along with their relative $\chi^2$ tests and Bayes factors $\ln \mathcal{B}_{ij}=\ln Z_i-\ln Z_\mathrm{\Lambda CDM}$ with respect to the $\Lambda$CDM model. Both the $w_0w_a$CDM and NMCQ models have shown a smaller $\chi^2$ test and moderate evidence $\ln \mathcal{B}=+3.69,\,+2.66$ over the $\mathrm{\Lambda}$CDM model, respectively, though with a slightly stronger preference for the $w_0w_a$CDM model due to the reduced constraining power in reconciling different datasets as shown below. Intriguingly, there appears to be more than $3\sigma$ evidence for the existence of a nonvanishing DM-$\varphi$ coupling with positive $n$ and $\beta$, as also shown in the appendixes, even updated with recent DES-Dovekie SN recalibration~\cite{DES:2025sig,Li:2026xaz}.

\begin{figure}[t]
  \includegraphics[width=\linewidth]{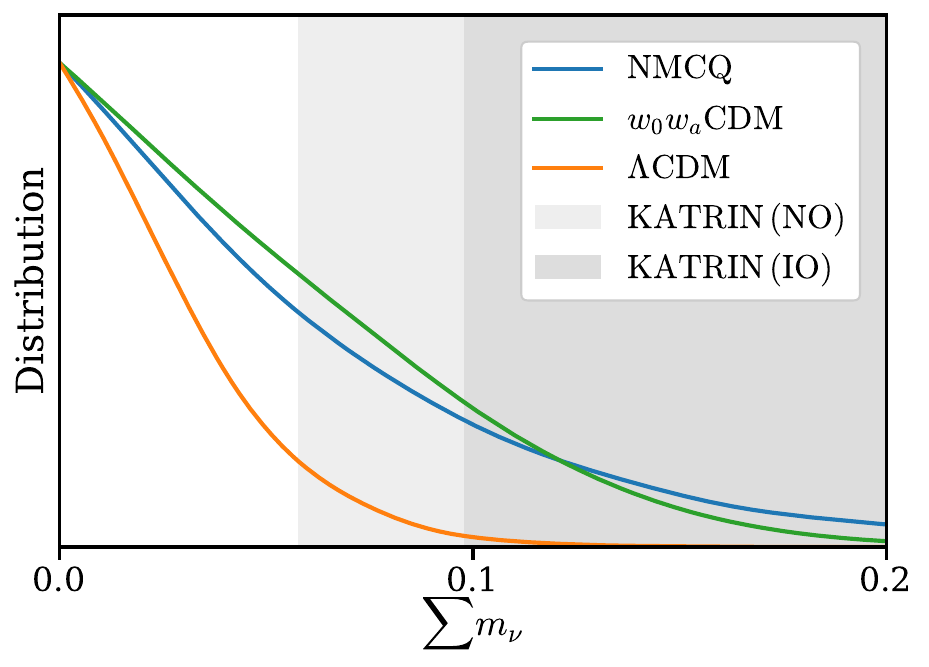}
  \caption{Cosmological constraints on neutrino mass in $\Lambda$CDM (orange), $w_0w_a$CDM (green), and NMCQ (blue) models from Planck-CMB+DESI BAO+DESY5 SNe, and the bounds from the KATRIN experiment for normal ordering (light gray) and inverted ordering (gray)~\cite{Esteban:2024eli}.} \label{fig:neutrino} 
\end{figure}

As for the $\Omega_\mathrm{m}$ tension, our NMCQ model gives rise to a $\Omega_\mathrm{m}$ value closer to the $\Lambda$CDM one than the $w_0w_a$CDM one, as shown in Table~\ref{tab:cosmo_params}. In particular, as shown in Fig.~\ref{fig:omegam} for each dataset constraint on $\Omega_\mathrm{m}$ and $H_0r_d$, the $\Omega_\mathrm{m}$ distribution is much more concentrated (overlapping within $1\sigma$) for our NMCQ model (solid) than the $w_0w_a$CDM model (dotted), thus largely resolving the $\Omega_\mathrm{m}$ tension. 

\begin{figure}[t]
  \centering 
  \includegraphics[width=\linewidth]{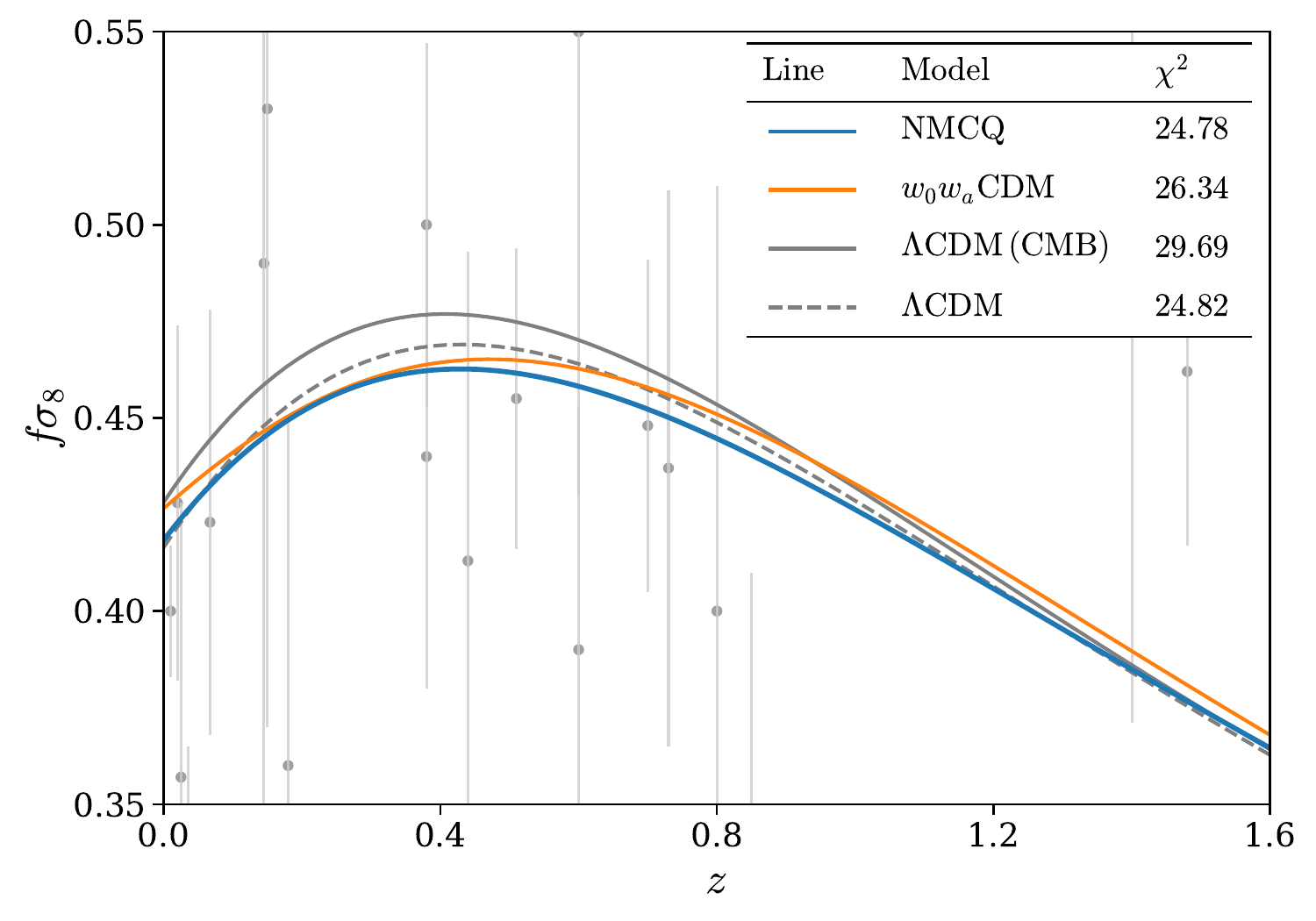}
  \caption{The theoretical predictions on $f\sigma_8$ from NMCQ (blue), $w_0w_a$CDM (orange), and $\Lambda$CDM (gray line), along with observational measurements (gray points). The parameters for each model's prediction are the best-fit constraints from CMB+BAO+SN, except for $\Lambda$CDM (CMB), which uses CMB only. The inset table presents the $\chi^2$ tests obtained with respect to the $f\sigma_8$ data alone for each model.}
  \label{fig:RSD} 
\end{figure}

We further find the other two discrepancies relieved as a result of the resolved $\Omega_\mathrm{m}$ tension. First, the neutrino-mass upper bound in the $\Lambda$CDM model is generally in tension with lower bounds from particle-physics experiments, while in our NMCQ model, the $95\%$ upper bound increases to $\sum  m_\nu < 0.179$ eV, comparable to that of the $w_0w_a$CDM model as shown in Fig.~\ref{fig:neutrino}. Second, the ``$\gamma$-tension''~\cite{Nguyen:2023fip} that the $\Lambda$CDM predicts a faster matter growth rate than $f\sigma_8$ measurements is also alleviated as shown in Fig.~\ref{fig:RSD}, where the NMCQ model exhibits the lowest matter growth rate with $\Delta\chi^2_{f\sigma_8}=-4.91$ reduction relative to $\Lambda\mathrm{CMB}\, (\mathrm{CMB})$ and $\Delta\chi^2_{f\sigma_8}=-1.56$ reduction to $w_0w_a$CDM, indicating improved agreement at the perturbation level. In addition, other cosmological results, such as $H_0$ and $S_8$, are not worsened in NMCQ model.


\section{Apparent phantom crossing}
 
For our unified fluid of coupled DM-$\varphi$ components, we can separate out a would-be standard CDM component from the apparent DM component, and then merge the rest into the quintessence field as the usual effective DE, that is,
\begin{align}
    \Delta\rho_\mathrm{DM}&=\rho_\mathrm{DM}-\rho_{\mathrm{DM},0}a^{-3}\label{Eq.Darksector}\\
    \rho_\mathrm{DE}^\mathrm{eff}&=\rho_\varphi+\Delta\rho_\mathrm{DM} \label{Eq.Darksector2},
\end{align}
where the noncold DM $\Delta\rho_\mathrm{DM}$ is the difference between the apparent DM and would-be standard CDM, and this definition is automatically subjected to a naive absence of noncold DM today, $\Delta\rho_{\mathrm{DM},0}=0$. The EOS of the effective DE can be obtained analytically as
\begin{equation}\label{Eq.wDE}
w_\mathrm{DE}^\mathrm{eff}=\frac{w_\varphi}{1+\Delta\rho_\mathrm{DM}/\rho_\varphi}.
\end{equation}
Different from recent realization (e.g., Ref.~\cite{Chakraborty:2025syu} and earlier Ref.~\cite{Das:2005yj}) of phantom crossing with help of an increasing function $\mathcal{A}(\varphi)$ of $\varphi$ so that $\Delta\rho_\mathrm{DM}=(\mathcal{A}-\mathcal{A}_0)\rho_\mathrm{CDM}$ can be negative due to $\mathcal{A}(\varphi)<\mathcal{A}(\varphi_0)$, our dilaton coupling $\mathcal{A}(\varphi)$ is a decreasing function of $\varphi$ so that $\Delta\rho_\mathrm{DM}$ is always positive and hence $w_\mathrm{DE}\geq w_\varphi>-1$. It seems the EOS of the usual effective DE in our model is always larger than the EOS of the quintessence field.

Is this in contradiction with the crossing behavior in $w_0w_a$CDM? To clarify this, we recall that the would-be standard CDM $\rho_\mathrm{DM,fid}\,a^{-3}$ is separated out only for comparison with $w_0w_a$CDM, and the fiducial value $\rho_\mathrm{DM,fid}$ is chosen as $\rho_{\mathrm{DM},0}$ in Eq.~\eqref{Eq.Darksector} only to expect $\Delta\rho_{\mathrm{DM},0}=0$ today. However,  there is currently no evidence to claim that all DM today are cold, $\Delta\rho_\mathrm{DM,0}=0$, and hence the fiducial $\rho_\mathrm{DM,fid}$ in the would-be standard CDM $\rho_\mathrm{DM,fid}\,a^{-3}$ should depend on the model to be compared with, rather than a parameter of NMCQ itself. As the observed crossing behavior is essentially a phenomenon raised by the CPL parameterization, the $\rho_\mathrm{DM,ref}$ in use should match the CDM component in the $w_0w_a$CDM model, yielding an apparent DE seen by $w_0w_a$CDM as
\begin{align}\label{Eq: pesdo energy density}
    \rho_\mathrm{DE}^\mathrm{app}&\equiv \left(\rho_\varphi+\rho_\mathrm{DM}\right)_\mathrm{NMCQ}-(\rho_\mathrm{CDM})_{w_0w_a\mathrm{CDM}}\\
    &=\rho_\mathrm{DE}^\mathrm{eff}+\rho_\mathrm{DM,0}a^{-3}-\rho_\mathrm{CDM,0}^\mathrm{CPL} a^{-3},
\end{align} 
whose EOS now reads
\begin{align}
w_\mathrm{DE}^\mathrm{app}=\frac{w_\mathrm{DE}^\mathrm{eff}}{1+(\rho_\mathrm{DM,0}-\rho_\mathrm{CDM,0}^\mathrm{CPL})a^{-3}/\rho_\mathrm{DE}^\mathrm{eff}}.
\end{align}
Since $\rho_\mathrm{DM,0}=\mathcal{A}_0\rho_\mathrm{CDM,0}$ does not necessarily equal to $\rho_\mathrm{CDM,0}^\mathrm{CPL}$, the denominator in the above apparent EOS could be negative, hence $w_\mathrm{DE}^\mathrm{app}$ could cross $-1$ with decreasing $\rho_\mathrm{DE}^\mathrm{eff}$ even if $w_\mathrm{DE}^\mathrm{eff}\geq w_\varphi>-1$ in our NMCQ model. See the \textit{Supplemental Appendices} for more details.

\begin{figure}[t]
  \includegraphics[width=\linewidth]{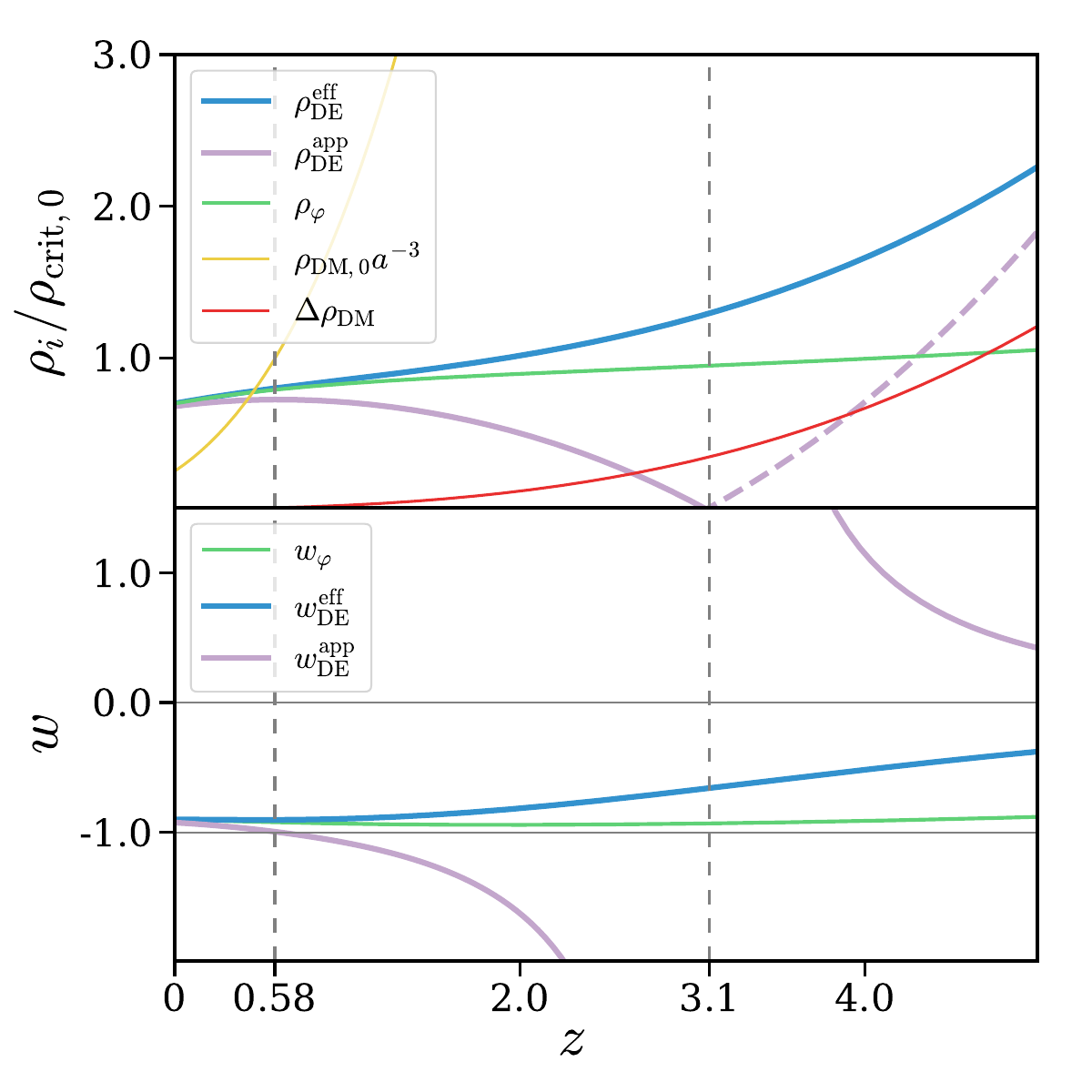}
  \caption{Energy densities and EOS parameters for the effective DE $\rho_\mathrm{DE}^\mathrm{eff}$, the apparent DE $\rho_\mathrm{DE}^\mathrm{app}$ (seen by $w_0w_a$CDM model), the quintessence field $\rho_\varphi$, the would-be CDM part $\rho_{\mathrm{DM},0}a^{-3}$, and the noncold DM part $\Delta\rho_\mathrm{DM}$. The dashed curve for the apparent DE presents its negative value, and vertical lines correspond to the redshifts where $w_\mathrm{DE}^\mathrm{app}=-1$ and $\rho_\mathrm{DE}^\mathrm{eff}=0$.}
  \label{fig:w}
\end{figure}

In Fig.~\ref{fig:w}, we present the energy-density $\rho_i$ and its EOS $w_i=-\dot{\rho}_i/(3H\rho_i)-1$ evolutions for all physical or artificial components using best-fit values of the $w_0w_a$CDM and NMCQ models in Table~\ref{tab:cosmo_params}. It is evident that the effective DE defined in Eq.~\eqref{Eq.Darksector2} does not exhibit any abnormal growth during expansion and never displays a crossing behavior. However, it is the apparent DE seen by the $w_0w_a$CDM model that changes the sign in its time derivative term $\dot{\rho}_\mathrm{DE}^\mathrm{app}$, and hence crosses $w=-1$ at $z=0.58$, consistent with what DESI found for phantom crossing around $z\sim0.5$ with CPL parameterization~\cite{DESI:2025zgx,DESI:2025fii}.

It is worth noting that data analysis does not depend on how we decompose the total dark sector $\rho_\varphi+\rho_\mathrm{DM}=(\rho_\varphi+\rho_\mathrm{DM}-\rho_\mathrm{DM,fid}a^{-3})+(\rho_\mathrm{DM,fid}a^{-3})$ into some DE part and CDM part, as we directly evolve $\rho_\varphi$ and $\rho_\mathrm{DM}$ in parameter sampling, and the mismatched DM part  $\Delta\rho_\mathrm{DM}\equiv\rho_\mathrm{DM}-\rho_\mathrm{DM,fid}a^{-3}$ with some fiducial choice on $\rho_\mathrm{DM,fid}$ only participates in the analysis of the crossing behavior of apparent DE EOS when a specific DE parameterization model is used. Accordingly, this apparent DE does not correspond to any real cosmological component, its crossing behavior is merely a modeling effect arising from attributing the mismatched term  $\rho_\mathrm{DM,NMCQ}-\rho_{\mathrm{CDM},w_0w_a\mathrm{CDM}}$ from the DM to the DE components. Therefore, the divergence in the apparent EOS $w_\mathrm{DE}^\mathrm{app}$ around $z\simeq 3.1$ and the negative energy density above that redshift do not reflect any theoretical crisis.

\section{Conclusions and discussions}\label{sec:condis}
The larger and more efficient survey from DESIY3 observations of BAO has claimed in their DR2 preliminary evidence for DDE with a crossing behavior. Although both Planck-CMB and DESY5-SNe admit some discrepancies or even tensions with DESI-BAO in both $\Lambda$CDM and $w_0w_a$CDM models when the matter fraction $\Omega_\mathrm{m}$ is specifically concerned, the DESI-BAO data alone still prefer a crossing behavior. In this paper, we adopt a string-theory-motivated quintessence field with the Ratra-Peebles potential and a dilaton coupling to the DM sector. We have detected more than $3\sigma$ evidence for such a DM-DE coupling. We have also derived an apparent crossing behavior when this model is misinterpreted as a $w_0w_a$CDM model. Moreover, unlike the $w_0w_a$CDM model that admits dispersive $\Omega_\mathrm{m}$ distributions for Planck, DESI, and DESY5, separately, our model admits much more concentrated $\Omega_\mathrm{m}$ constraints without tensions. Several discussions follow as below:

First, the DM-DE interaction in our model is free of current fifth-force constraints~\cite{Carroll:2008ub,Bai:2009it,Carroll:2009dw} and requires no screening mechanism at local scales, as the DM-DE interaction is actually subject to the dark force, whose constraint from tidal tails on $\beta<0.7$~\cite{Kesden:2006vz} is well above our best-fit value of $\beta\sim 0.05$. The unified dark fluid from the DM-DE interaction makes it subtle to separate one from the other, and our study suggests that it remains of great theoretical interest to explore the unified dark-fluid scenarios~\cite{Wang:2024rus,Kamenshchik:2001cp,Bilic:2001cg,Bento:2002ps,Makler:2002jv,Sandvik:2002jz,Scherrer:2004au,Zhang:2004gc,Cai:2015rns,Koutsoumbas:2017fxp,Ferreira:2018wup}, especially beyond the general relativity framework.


Second, it has been recently shown in Ref.~\cite{Lewis:2024cqj} that the null energy condition can rule out certain regions supported by some BAO distance scales for any physical noninteracting DE model within FLRW cosmology. Intriguingly, the regions in tension with the $\Lambda$CDM model from current DESI BAO data arise primarily in the directions breaking the null-energy condition, thus unless FLRW cosmology is broken~\cite{Colgain:2024mtg}, one has to consider either the interacting DE model (or equivalently noncold dynamical DM~\cite{Yang:2025ume,Wang:2025zri,Kumar:2025etf,Abedin:2025dis,Li:2025eqh}) or the broken null-energy condition (for example, the quintom model~\cite{Feng:2004ad,Feng:2004ff,Guo:2004fq}), and even both. This goes along with findings from Ref.~\cite{Ye:2025ark}. 

Third, this study only considers a positive prior for the coefficient $\beta$ in the exponent of the dilaton coupling $\mathcal{A}(\varphi)=e^{-\beta\varphi/M_\mathrm{Pl}}$. A negative $\beta$ could also mimic the crossing behavior but correspond to rather different dynamics---the chameleon DE~\cite{Cai:2021wgv}---that resolves the Hubble tension~\cite{Bernal:2016gxb,Verde:2019ivm,Riess:2020sih,Abdalla:2022yfr,Hu:2023jqc,Vagnozzi:2023nrq,Cai:2023sli}, not at the background level but at the perturbation level: overdensity regions would admit higher effective potential minima thus expand locally faster than the background, as also confirmed recently with the data~\cite{Yu:2022wvg}. Regions where SNe, Cepheids, and Milky Way are located with only $6\%$ overdensity just below the homogeneity scale are enough to contribute $6$ km/s/Mpc in total on top of background expansion to fill in the Hubble tension. A full analysis will be reported.

\begin{acknowledgments}
We are grateful to Yun-He Li, Gen Ye, and Meng-Xiang Lin for insightful discussions, as well as Zheng Cheng and Mengjiao Lyu for computational support. This work is supported by the National Key Research and Development Program of China Grants No. 2021YFC2203004, No. 2021YFA0718304, and No.2020YFC2201501, the National Natural Science Foundation of China Grants No. 12422502, No. 12547110, No.12588101, No. 12235019, and No. 12447101; and the China Manned Space Program Grant No. CMS-CSST-2025-A01. We also acknowledge the use of the HPC Cluster of ITP-CAS.
\end{acknowledgments}

\section*{DATA AVAILABILITY}
The data that support the findings of this article are not publicly available because they are owned by a third party and the terms of use prevent public distribution. The data are available from the authors upon reasonable request.


\appendix

\section{INITIAL CONDITION}

Given the runaway form of the effective potential, the initial condition of $\varphi$ may become significant. Since a scalar field with a power-law potential typically exhibits scaling behavior in the early Universe, we adopted the attractor solution starting from $z\simeq 10^{12}$ as a common approximation~\cite{Copeland:1997et}:
\begin{equation}
  \varphi_\mathrm{r} (a)=\varphi_{\mathrm{r},0} a^\lambda,
\end{equation}
where we denote the initial value of $\varphi$ as $\varphi_\mathrm{r}$ deep into the radiation era. Substituting $\varphi_\mathrm{r}$ into the EOM for $\varphi$ and neglecting the coupling term, one can derive $\lambda$ and $\varphi_\mathrm{r}$ as
\begin{align}
  \lambda&=\frac{4}{n+2},\label{Eq.lambda}\\
  \varphi_\mathrm{r}&=\left(\frac{\alpha n (n+2)^2}{4(6+n)H^2}\right)^\frac{1}{2+n},\label{Eq.phiini}\\
  \dot{\varphi}_\mathrm{r}&=\lambda H\varphi_\mathrm{r}.\label{Eq.phidotini}
\end{align}
This solution will be used as the initial condition for solving the EOM of $\varphi$.

Although the attractor solution is commonly used for the inverse power-law scalar fields at early times, its application to our NMCQ model raises two significant concerns. First, a key premise that the coupling term is negligible lacks justification. Second, to physically treat $\varphi_\mathrm{r}$ and $\dot{\varphi}_\mathrm{r}$ as fixed initial conditions rather than sampling parameters, we need to illustrate that the effect of $\varphi_\mathrm{r}$ on the solution is quite weak.

\begin{figure}[h]
  \includegraphics[width=\linewidth]{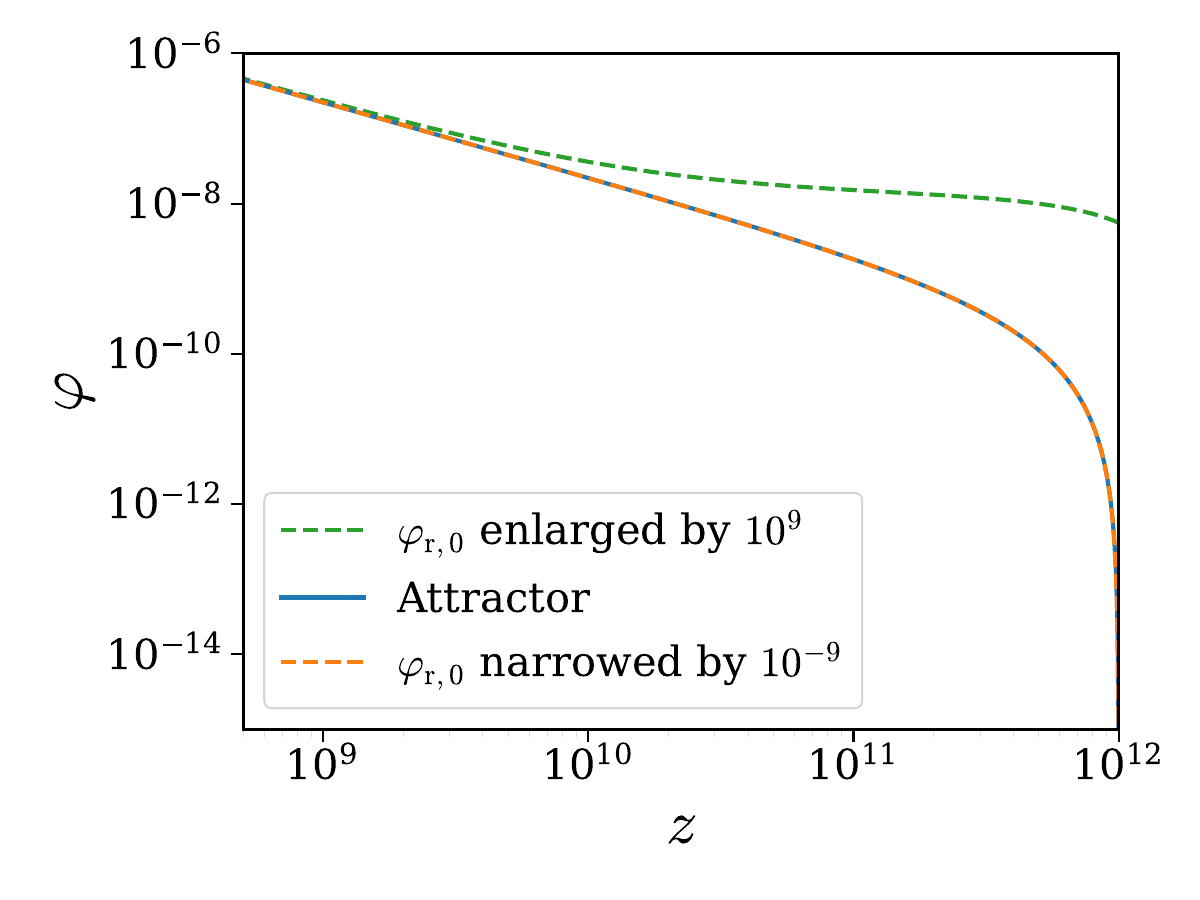}
  \caption{Evolution of $\varphi$ during $10^9\leq z<10^{12}$, where the solutions with narrowed and enlarged $\varphi_{\mathrm{r},0}$ are separately presented with orange and green dashed lines, and the blue solid line is for the attractor solution. The overlap of the blue and orange curves arises from the quite rapid increase of $\varphi$ in the very early Universe.}\label{Fig.phi} 
\end{figure}

To address these concerns, we both increase and suppress the initial values by a factor of $10^9$, and plotted the evolution of $\varphi$ in Fig.~\ref{Fig.phi}. Notably, the dynamics of $\varphi$ are almost identical after $z=10^9$, and the shooting parameters $\alpha$ and $\varphi_0$ varied by less than $10^{-7}$ under these three scenarios. As for the effects of the coupling term, we note quintessence decays as $\rho_\varphi \propto a^{-1.5}$ during radiation domination based on Eq.~\eqref{Eq.lambda}, while the DM decays as $a^{-3}$, and the coupling term will become significant as redshift increases. This suggests that the attractor solution most likely breaks down at high redshift. However, the results of $\varphi$ reveal that $V'(\varphi)$ exceeds $\beta \rho_\mathrm{DM}$ by at least one order of magnitude across all redshifts. As illustrated in Fig.~\ref{Fig.phi}, the value of the scalar field will increase rapidly or freeze until $V'(\varphi) \gtrsim \beta\rho_{\mathrm{DM}}$ to restore its scaling behavior and return to the attractor. 
This justifies neglecting the coupling term in our initial approximation. Therefore, we can safely set the attractor as a fixed physical initial condition, as long as $\varphi_\mathrm{r}$ and $\dot{\varphi}_\mathrm{r}$ are not too large to thaw before $z_\mathrm{eq}$.

\section{MATCHING CONDITION}

It should be noted that the two coefficients, $\alpha$ and $\mathcal{A}_0$, should coincide with the solutions derived from them. Here we will first investigate the coefficient of the coupling term, $\mathcal{A}_0$. As shown in Eq.~\eqref{Eq.rhoDM}, all denominators represent physical quantities evaluated at a fixed time, which is conventionally taken as the present epoch ($a_0=1$) with $\rho_{\mathrm{DM},0} = \Omega_\mathrm{DM}\rho_{\mathrm{crit},0}$. Consequently, $\mathcal{A}_0$ essentially encodes the current field value by
\begin{equation}
  \mathcal{A}_0=\exp \left(-\beta\frac{\varphi_0}{M_\mathrm{Pl}}\right),
\end{equation}
then the DM density reads
\begin{equation}
  \rho_\mathrm{DM}=\rho_{\mathrm{DM},0}a^{-3}\exp \left(-\frac{\beta(\varphi -\varphi_0)}{M_\mathrm{Pl}}\right).
\end{equation}

As an intrinsic component of the solution, $\varphi_0$ directly influences the coefficients in its EOM. The input parameter $\varphi_0$ must mathematically equal the solved field value $\varphi(z=0)$ at the present day. This constraint implies $\varphi_0$ cannot be treated as a free parameter unless we can start solving the EOM at $z=0$. However, $\varphi_0$ becomes essentially immutable once the parameters are fixed, as illustrated in the last section. This prevents us from arbitrarily specifying $\varphi_0$ and $\dot{\varphi}_0$ as external priors.

In addition, the constraint on $\alpha$ is a physical premise for energy density via $\Omega_\mathrm{DE}\approx1-\Omega_\mathrm{DM}$ at late times. The Hubble parameter used in Eqs.~\eqref{Eq.EOMphi} and~\eqref{Eq.EOMDM} should be consistent with the input parameter $H_0$ by
\begin{equation}
  \frac{\rho_{\mathrm{DM},0}+\frac{1}{2}\dot{\varphi_0}^2+V(\varphi_0)}{3 M_\mathrm{Pl}^2 H_0^2}\approx1,
\end{equation}
where we have ignored the radiation at low redshift. This requires $\alpha$ to be determined by the current critical energy density, $\rho_{\mathrm{crit},0}$. As an estimation based on the energy scale, the potential of quintessence should approximately approach the energy density today, as follows:
\begin{equation}
  \alpha \Lambda^4\approxeq 3M_\mathrm{Pl}^2 H_0^2.
\end{equation}
Hence, we simply set $\Lambda^4\equiv3M_\mathrm{Pl}^2 H_0^2$ in $V(\varphi)$ and shoot for the values of $\varphi_0$ and $\alpha$ of order $\mathcal{O}(1)$, similar to Ref.~\cite{Cai:2021wgv}. 

To technically determine the correct matching conditions for $(\alpha, \varphi_0)$, the Broyden iteration method can be performed as below~\cite{Broyden:1965isd, gay1979some}. 
We define a two-dimensional residual function $\mathbf{f}(\mathbf{x})$ whose components quantify the mismatch between the evolved quantities and their target values at $a=1$. The vector $\mathbf{x}$ contains the initial guesses for $(\alpha, \varphi_0)$, for example, $(1,0.5)$ (other $\mathcal{O}(1)$ values are also allowed). 
At each iteration, the update is computed by
\begin{equation}
\mathbf{x}_{k+1} = \mathbf{x}_k - \mathbf{H}_k \cdot \mathbf{f}_k,
\end{equation}
where $k$ denotes the iterations and $\mathbf{H}_k$ is the approximate inverse Jacobian matrix calculated by the Broyden rank-one formula,
\begin{equation}
\mathbf{H}_{k+1} = \mathbf{H}_k + 
\frac{(\Delta \mathbf{x}_k - \mathbf{H}_k \Delta \mathbf{f}_k) \otimes \Delta \mathbf{x}_k}
{\Delta \mathbf{x}_k^\intercal \cdot \Delta \mathbf{f}_k},
\end{equation}
with
\begin{align}
\Delta \mathbf{x}_k &= \mathbf{x}_{k+1} - \mathbf{x}_k, \\
\Delta \mathbf{f}_k &= \mathbf{f}_{k+1} - \mathbf{f}_k.
\end{align}
The iteration continues until the norm of the residual is satisfied, as follows:
\begin{equation}
\| \mathbf{f}(\mathbf{x}) \| < \varepsilon.
\end{equation}
In this work, $\varepsilon$ was set as $10^{-5}$ for all calculations.

\section{PRIORS AND POSTERIORS}

\newcommand{\pA}[1]{\parbox[t]{0.10\linewidth}{\raggedright #1}}
\newcommand{\pB}[1]{\parbox[t]{0.15\linewidth}{\raggedright #1}}
\newcommand{\pC}[1]{\parbox[t]{0.20\linewidth}{\raggedright #1}}
\newcommand{\pD}[1]{\parbox[t]{0.24\linewidth}{\raggedright #1}}
\newcommand{\pE}[1]{\parbox[t]{0.24\linewidth}{\raggedright #1}}

\begin{table*}[t]
  \centering
  \caption{Priors for all model and cosmological parameters. The last three columns list the priors for nested sampling, MCMC sampling using CMB, and MCMC sampling without using CMB, respectively. $\mathcal{N}$ and $\mathcal{U}$ denote Gaussian and flat priors, while $\delta$ corresponds to a fixed value. All cosmological parameters taken in nested sampling are the same.}\label{Tab:prior}
  \renewcommand{\arraystretch}{1.4}
  \begin{tabular}{lllll}
  \hline\hline
  \pA{\textbf{Model}} & \pB{\textbf{Parameter}} & \pC{\textbf{Nested Sampling}} & \pD{\textbf{MCMC with CMB}} & \pE{\textbf{MCMC without CMB}} \\
  \hline
  \multirow{6}{*}{$\Lambda$CDM} 
  & $\Omega_b h^2$ & $\mathcal{U}[0.021,\ 0.024]$ & $\mathcal{U}[0.005,\ 0.1]$ & $\mathcal{N}[0.02237,\ 0.00015]$ \\
  & $\Omega_c h^2$ & $\mathcal{U}[0.10,\ 0.13]$ & $\mathcal{U}[0.001,\ 0.99]$ & $\mathcal{U}[0.001,\ 0.99]$ \\
  & $H_0$  & $\mathcal{U}[61,\ 75]$ & $\mathcal{U}[20,\ 100]$ & $\mathcal{U}[45,\ 90]$ \\
  & $\tau$ & $\mathcal{U}[0.02,\ 0.2]$ & $\mathcal{U}[0.01,\ 0.8]$ & $\delta[0.055]$ \\
  & $\log(10^{10} A_s)$ & $\mathcal{U}[2.9,\ 3.2]$ & $\mathcal{U}[1.61,\ 1.91]$ & $\delta [3.045]$ \\
  & $n_s$ & $\mathcal{U}[0.93,\ 1.00]$ & $\mathcal{U}[0.8,\ 1.2]$ & $\delta[0.9649]$ \\
  \hline
  $w_0w_a$CDM & $w_0$ & $\mathcal{U}[-2.5,\ 1.5]$ & $\mathcal{U}[-50,\ 20]$ & $\mathcal{U}[-150,\ 20]$ \\
      & $w_a$ & $\mathcal{U}[-3.5,\ 1]$ & $\mathcal{U}[-3,\ 2]$ & $\mathcal{U}[-50,\ 20]$ \\
  \hline
  NMCQ & $n$ & $\mathcal{U}[0.01,\ 2.0]$ & $\mathcal{U}[0.01,\ 4.0]$ & $\mathcal{U}[0.01,\ 2.0]$ \\
      & $\beta$ & $\mathcal{U}[0,\ 0.2]$ & $\mathcal{U}[0,\ 0.5]$ & $\mathcal{U}[0,\ 0.3]$ \\
  \hline\hline
  \end{tabular}
\end{table*}

For model comparison, we employ Bayesian analysis based on the relative Bayes factor in logarithmic space, $\ln \mathcal{B}_{ij}=\ln Z_i-\ln Z_\mathrm{\Lambda CDM}$~\cite{Rigault:2014kaa, Handley:2015vkr}. We use the revised Jeffrey's scale~\cite{Jeffreys:1939xee} to interpret the results. The Bayes evidence was calculated by nested sampling using the public package~\texttt{PolyChord}~\cite{Handley:2015fda,Handley:2015vkr}. The sampling was completed while the evidence contained in live points was less than $\Delta \ln Z=0.001$. To determine the constraint on external prior and obtain a more accurate posterior, MCMC analyses were also performed using the \texttt{mcmc} module of \texttt{Cobaya}~\cite{Lewis:2002ah, Lewis:2013hha, Neal:2005uqf}, where the final Gelman-Rubin diagnostic of MCMC sampling was limited to $R-1<0.01$~\cite{Gelman:1992zz}. To analyze and plot the MCMC results, we used the public package \texttt{Getdist}~\cite{Lewis:2019xzd}.

The external priors for different models, sampling methods, and likelihoods are listed in Table~\ref{Tab:prior}. 
It is necessary to explain the flat priors in the third column of Table~\ref{Tab:prior} since we adopt narrower parameter ranges compared to the conventional prior used in $\mathrm{\Lambda CDM}$ as shown in the fourth column. The shooting method is employed for solving EOMs to ensure the consistency between the initial condition and the resulting solution. However, this may fail under unphysical parameter combinations, for example, an oversize $\Omega_\mathrm{m}=0.99$. Since nested sampling explores the entire prior space, we need to restrict priors to physically viable regions to prevent such failures, similar to Ref.~\cite{Ye:2024zpk}.
To demonstrate its validity and reduce the deviation of Bayes factors raised by this, we unified the external prior for each parameter in all models, and used both MCMC and nested sampling to calculate the posterior and Bayes evidence, respectively. As a result, the external prior for nested sampling can still cover the $5\sigma$ range for all parameters.

\begin{figure*}[!ht]
  \includegraphics[width=1.\linewidth]{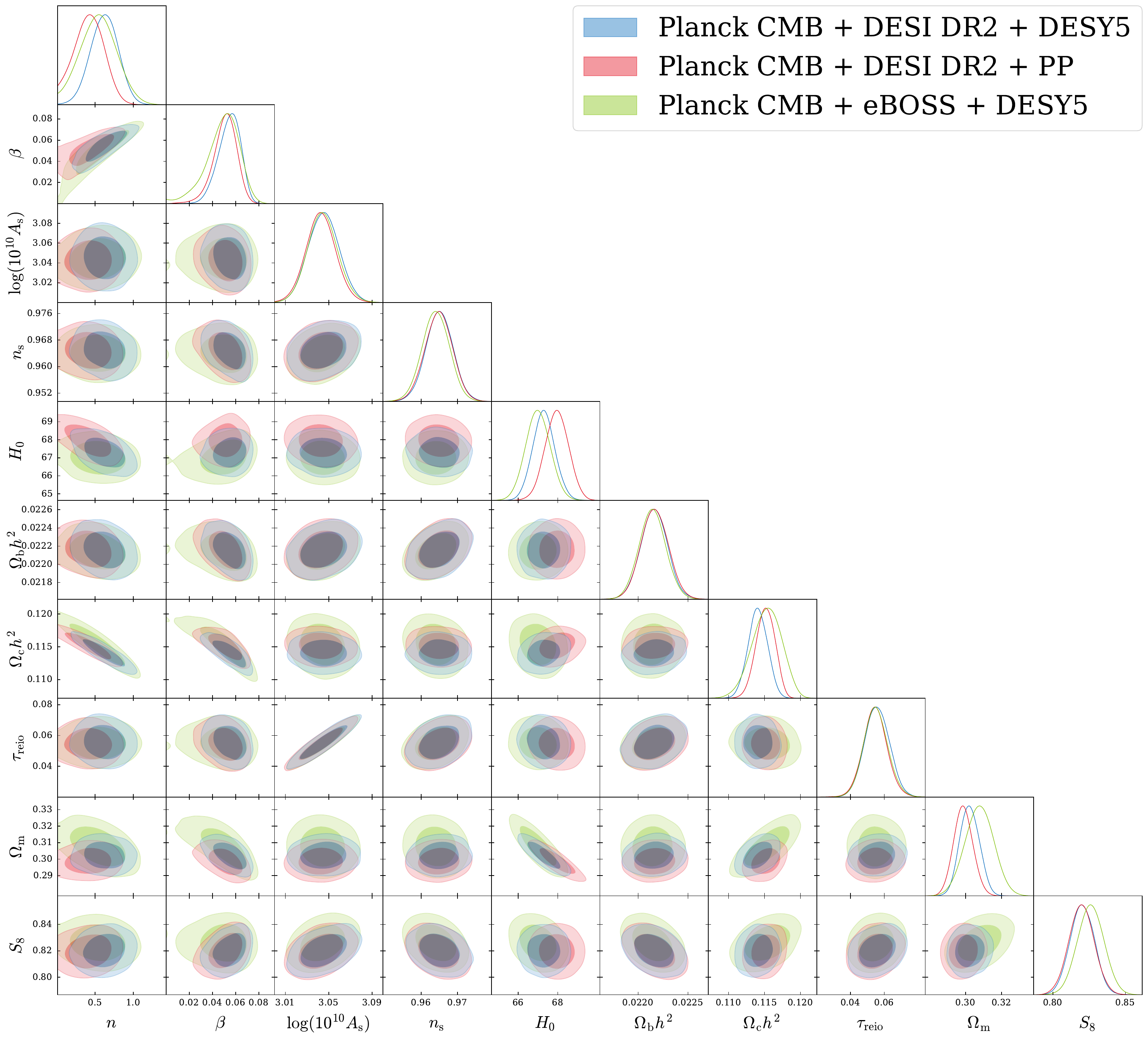}
  \caption{\label{fig.fullMCMC} Full posterior distribution of all cosmological parameters in our NMCQ model.}
\end{figure*}

To compare the preferences of different datasets for NMCQ, we also used the SNe Ia datasets of the PantheonPlus sample, denoted as PP~\cite{Brout:2022vxf}, and the DR16 BAO measurements by the extended Baryon Oscillation Spectroscopic Survey, denoted as eBOSS~\cite{eBOSS:2020yzd}. 
All of the parameter distributions in NMCQ are shown in Fig.~\ref{fig.fullMCMC}. For all combinations of datasets, the evidence of a nonvanishing coupling with positive $\beta$ is over $2\sigma$. Compared to Planck-CMB+DESI DR2+DESY5, the other two combinations of datasets prefer a smaller $n$, while it is still nonzero at about $2\sigma$. This suggests that both DESI DR2 and DESY5 can provide evidence for the existence of the nonminimally coupled quintessence rather than a cosmological constant $\Lambda$ alone.

\section{Full redshift evolutions}

\textbf{Effective DE:} In general, the DE-DM interaction introduces an energy density flow $Q=[\mathcal{A}'(\varphi)/\mathcal{A}(\varphi)]\dot{\varphi}\rho_\mathrm{DM}$ between the apparent DM and quintessence field via their EOMs,
\begin{align}
\dot{\rho}_\varphi+3H(1+w_\varphi)\rho_\varphi&=-Q,\\
\dot{\rho}_\mathrm{DM}+3H\rho_\mathrm{DM}&=+Q.
\end{align}
Due to the DE-DM interaction, the apparent DM sector does not evolve exactly as the standard CDM. Thus, one usually separates out a would-be standard CDM part $\rho_\mathrm{DM,fid}a^{-3}\equiv\rho_\mathrm{DM}^\mathrm{fid}$ for some fiducial value $\rho_\mathrm{DM,fid}$, and then defines the noncold DM part as
\begin{align}
\Delta\rho_\mathrm{DM}\equiv\rho_\mathrm{DM}-\rho_\mathrm{DM}^\mathrm{fid},
\end{align}
which can be absorbed into the quintessence part to define the so-called effective DE sector,
\begin{align}
\rho_\mathrm{DE}^\mathrm{eff}=\rho_\varphi+\Delta\rho_\mathrm{DM}.
\end{align}
Now the EOMs for each sector of the unified fluid decomposition $\rho_\varphi+\rho_\mathrm{DM}=\rho_\mathrm{DE}^\mathrm{eff}+\rho_\mathrm{DM}^\mathrm{fid}$ become
\begin{align}
\dot{\rho}_\mathrm{DE}^\mathrm{eff}+3H(1+w_\mathrm{DE}^\mathrm{eff})\rho_\mathrm{DE}^\mathrm{eff}&=0,\\
\dot{\rho}_\mathrm{DM}^\mathrm{fid}+3H\rho_\mathrm{DM}^\mathrm{fid}&=0,
\end{align}
where the EOS of the above effective DE is the usual one used in the interacting DE models (e.g., Chakraborty et al.~\cite{Chakraborty:2025syu}),
\begin{align}
w_\mathrm{DE}^\mathrm{eff}=\frac{w_\varphi}{1+(\rho_\mathrm{DM}-\rho_\mathrm{DM}^\mathrm{fid})/\rho_\varphi}=w_\varphi\frac{\rho_\varphi}{\rho_\mathrm{DE}^\mathrm{eff}}.
\end{align}
Note that, one can naively choose $\rho_\mathrm{DM}^\mathrm{fid}\equiv\rho_\mathrm{DM,fid}a^{-3}=\rho_\mathrm{DM,0}a^{-3}$ so that $\Delta\rho_\mathrm{DM,0}=\rho_\mathrm{DM,0}-\rho_\mathrm{DM,0}a_0^{-3}=0$, that is, all DM today is cold, which has not been verified yet. As long as $\rho_\mathrm{DM}-\rho_\mathrm{DM}^\mathrm{fid}$ once evolves to be negative, the effective EOS $w_\mathrm{DE}^\mathrm{eff}$ could cross the phantom divide, which is the traditional way to interpret the recent DESI result.

\textbf{Apparent DE:} Here, we provide an alternative viewpoint. If we want to use a specific DE parameterization model, say, the CPL model, to interpret the data, then the apparent DE seen by the CPL model should be
\begin{align}
\rho_\mathrm{DE}^\mathrm{app}=(\rho_\varphi+\rho_\mathrm{DM})-\rho_\mathrm{CDM}^\mathrm{CPL}=(\rho_\mathrm{DE}^\mathrm{eff}+\rho_\mathrm{DM}^\mathrm{fid})-\rho_\mathrm{CDM}^\mathrm{CPL},
\end{align}
and the EOMs for each sector of the unified fluid decomposition $\rho_\varphi+\rho_\mathrm{DM}=\rho_\mathrm{DE}^\mathrm{eff}+\rho_\mathrm{DM}^\mathrm{fid}=\rho_\mathrm{DE}^\mathrm{app}+\rho_\mathrm{CDM}^\mathrm{CPL}$ becomes
\begin{align}
\dot{\rho}_\mathrm{DE}^\mathrm{app}+3H(1+w_\mathrm{DE}^\mathrm{app})\rho_\mathrm{DE}^\mathrm{app}&=0,\\
\dot{\rho}_\mathrm{DM}^\mathrm{CPL}+3H\rho_\mathrm{DM}^\mathrm{CPL}&=0,
\end{align}
where the EOS of the apparent DE can be easily computed as
\begin{align}
w_\mathrm{DE}^\mathrm{app}&=\frac{w_\mathrm{DE}^\mathrm{eff}}{1+(\rho_\mathrm{DM}^\mathrm{fid}-\rho_\mathrm{CDM}^\mathrm{CPL})/\rho_\mathrm{DE}^\mathrm{eff}}=w_\mathrm{DE}^\mathrm{eff}\frac{\rho_\mathrm{DE}^\mathrm{eff}}{\rho_\mathrm{DE}^\mathrm{app}}\\
&=\frac{w_\varphi}{1+(\rho_\mathrm{DM}-\rho_\mathrm{CDM}^\mathrm{CPL})/\rho_\varphi}=w_\varphi\frac{\rho_\varphi}{\rho_\mathrm{DE}^\mathrm{app}}.
\end{align}
Note that the above relation $w_\varphi\rho_\varphi=w_\mathrm{DE}^\mathrm{eff}\rho_\mathrm{DE}^\mathrm{eff}=w_\mathrm{DE}^\mathrm{app}\rho_\mathrm{DE}^\mathrm{app}$ is physically intuitive as the would-be standard CDM part $\rho_\mathrm{DM}^\mathrm{fid}=\rho_\mathrm{DM,fid}a^{-3}$ or $\rho_\mathrm{CDM}^\mathrm{CPL}=\rho_\mathrm{CDM,0}^\mathrm{CPL}a^{-3}$ does not contribute to the pressure. 
Also note that the apparent EOS does not depend on the fiducial choice of the observationally unknown $\rho_\mathrm{DM,fid}$, which is one advantage over the effective EOS.
Again, as long as $\rho_\mathrm{DM}-\rho_\mathrm{DM}^\mathrm{CPL}$ once evolves to be negative, the apparent EOS $w_\mathrm{DE}^\mathrm{app}$ could cross the phantom divide to explain the DESI result. In specific, if we choose the fiducial value $\rho_\mathrm{DM,fid}=\rho_\mathrm{DM,0}$ from our NMCQ model, then $\rho_\mathrm{DM,0}=\mathcal{A}(\varphi_0)\rho_\mathrm{CDM,0}$ does not necessarily equal to $\rho_\mathrm{CDM,0}^\mathrm{CPL}$ since $A(\varphi_0)$ is not necessarily 1, and $\rho_\mathrm{CDM,0}$ in our NMCQ model does not necessarily equal to $\rho_\mathrm{CDM,0}^\mathrm{CPL}$ in CPL model. It is this matter fraction mismatch that causes the apparent phantom crossing behavior if the CPL model is adopted for interpretation.

The key argument here is that, for most of the interacting DE models on the market (e.g., Ref.~\cite{Chakraborty:2025syu}), the effective EOS $w_\mathrm{DE}^\mathrm{eff}$ (with $\rho_\mathrm{DM}^\mathrm{fid}=\rho_\mathrm{DM,0}a^{-3}$) roughly agrees with the apparent EOS $w_\mathrm{DE}^\mathrm{app}$ (independent of the fiducial value $\rho_\mathrm{DM,fid}$). However, there exists another branch of possibility that the effective EOS $w_\mathrm{DE}^\mathrm{eff}$ could be rather different from the apparent EOS $w_\mathrm{DE}^\mathrm{app}$, and our model is one such simple illustration example due to the mismatched matter density between the underlying model and CPL parameterization. The novelty of this difference is that the apparent EOS is more suitable to interpret the data than the effective EOS before we can precisely measure the present-day DM property (that is, how many DM today are exactly cold).

\begin{figure}[h]
  \includegraphics[width=\linewidth]{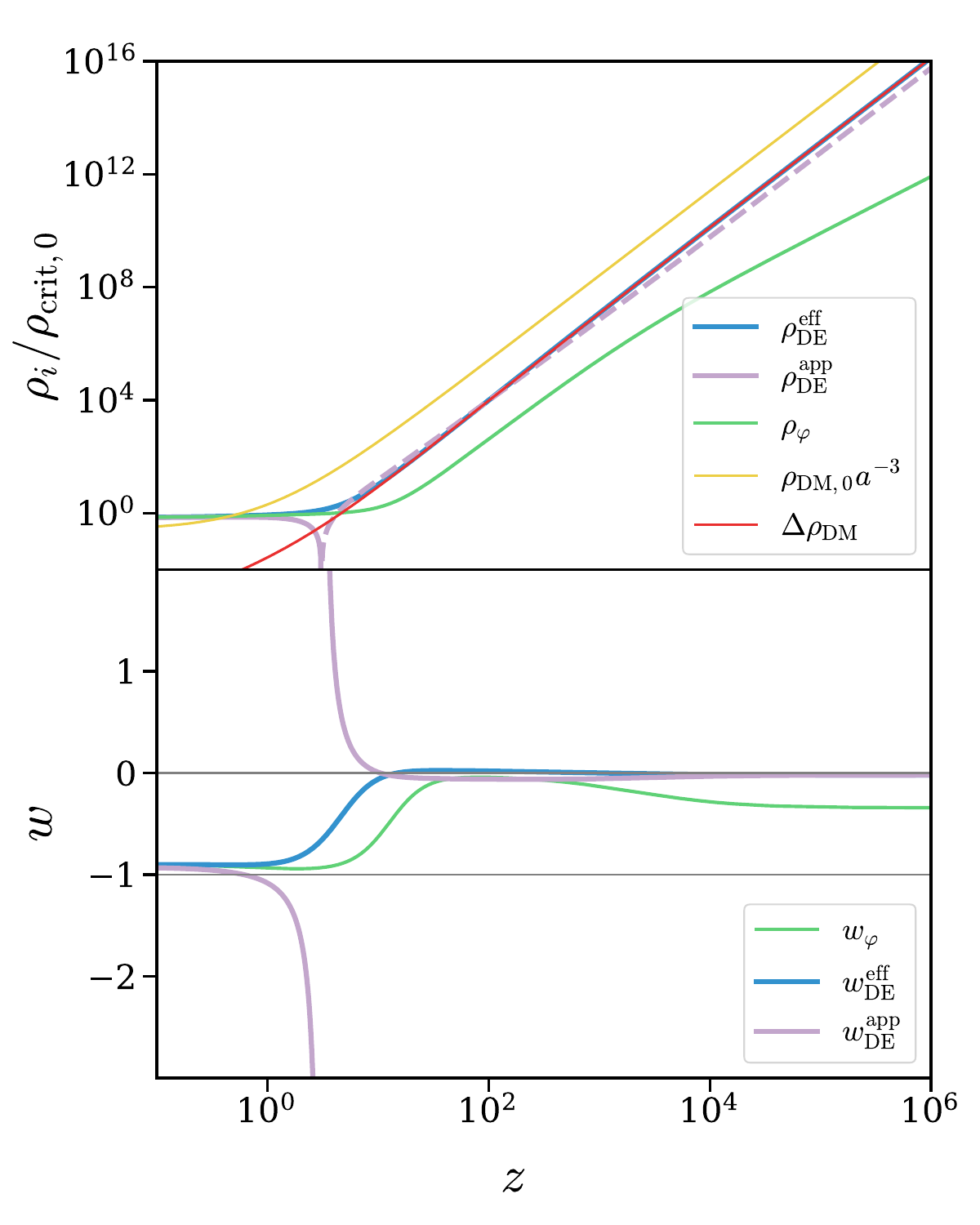}
  \caption{Full redshift evolutions of energy densities and EOS parameters for the apparent DE $\rho_\mathrm{DE}$, the observational DE $\rho_\mathrm{DE}^\mathrm{obs}$, the quintessence field $\rho_\varphi$, the would-be CDM $\rho_{\mathrm{DM},0}a^{-3}$, and the noncold DM part $\Delta\rho_\mathrm{DM}$. The dashed curve for the observational DE presents its negative value.}
  \label{fig:whiz}
\end{figure}

By the definition of observational EOS, $w_i=-\dot{\rho_i}/3H\rho_i-1$, the full redshift evolutions of the energy density and EOS parameters of our NMCQ model are presented in Fig.~\ref{fig:whiz}, based on which we can categorize cosmic history into three distinct phases:
\begin{enumerate}
    \item[i.] Quintessence dominates the Universe at late times ($0<z\leq 0.45$), and approximately freezes back to $z=8$. This behavior is similar to the cosmological constant $\Lambda$. Therefore, the EOS of the observational DE is increasing and larger than $-1$ at low redshifts. Before that, quintessence decays at a rate lower than DM at early times. 
    \item[ii.] DM takes the dominant place of $\varphi$ at $0.45<z\leq 3300$ due to the freeze of quintessence, and the difference between $\rho_\mathrm{DM,NMCQ}-\rho_{\mathrm{CDM},w_0w_a\mathrm{CDM}}$ can exceed $\rho_\varphi$ before $z=3.1$. 
    \item[iii.]  The redshift of the matter-radiation equality $z_\mathrm{eq}$ does not change significantly.
\end{enumerate}
These results demonstrate that the crossing behavior and energy dispersion of observational DE are fundamentally attributable to the cosmological transition from DM to DE dominance, and this exactly explains why DESI found an apparent phantom crossing around $z\sim0.5$. Furthermore, as shown in Fig.~\ref{fig:whiz}, the noncold DM part $\Delta \rho_\mathrm{DM}$ closely matches the deviation in the mismatched term $\rho_{\mathrm{DM},0}a^{-3}$ between NMCQ and Planck-$\Lambda$CDM models at high redshifts ($z>100$). This correspondence ensures recovery of the Planck-$\Lambda$CDM dark matter fraction at recombination, thereby preserving the integrity of CMB spectra.

\section{More on $\Omega_\mathrm{m}$ tension}

\begin{figure*}[t]
  \includegraphics[width=\linewidth]{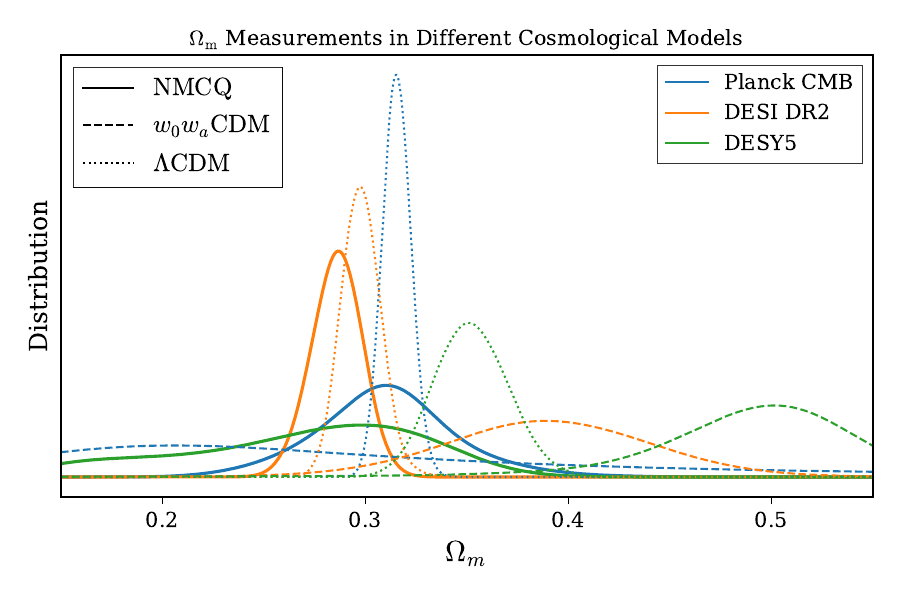}
  \caption{Cosmological constraints on the matter fraction parameter $\Omega_\mathrm{m}$ from Planck-CMB (blue), DESI DR2 (red), and DESY5 (green) in the $\Lambda$CDM (dotted), $w_0w_a$CDM (dashed), and NMCQ (solid) models.}
  \label{fig.omegaw0wa} 
\end{figure*}

The constraints on $\Omega_\mathrm{m}$ from Planck-CMB (blue), DESI DR2 (red), and DESY5 (green) in $\Lambda$CDM (dotted), $w_0w_a$CDM (dashed), and NMCQ (solid) are compared in Fig.~\ref{fig.omegaw0wa}. Compared to $\mathrm{\Lambda CDM}$, the CPL parametrization method indeed worsens the tensions among CMB, DESI, and DESY5, while our NMCQ model allows for more concentrated distributions within $1\sigma$ overlapping among them. The posterior space of each single dataset became quite large in $w_0w_a$CDM, and the best-fit $\Omega_\mathrm{m}$ of DESY5 was even larger than it was in $\mathrm{\Lambda CDM}$. Although $\Delta \chi ^2_{\mathrm{MAP}}$ can be significantly reduced in $w_0w_a$CDM, the discrepancy in the best-fit value of each dateset became even more significant, and the role of the dynamic of DE is more likely to weaken the ability of data to constrain the cosmological parameters. We summarize the $\Omega_\mathrm{m}$ constraints in Table~\ref{tab:omega_m}. Notably, only in the NMCQ model does the best-fit value from the combined dataset fall within the $2\sigma$ level inferred from each individual dataset. This perhaps implies that parametrization methods such as $w_0w_a$CDM may become less favored by precise observations in the future. 
\newcommand{\omA}[1]{\parbox[t]{0.28\linewidth}{\raggedright #1}}
\newcommand{\omB}[1]{\parbox[t]{0.12\linewidth}{\centering #1}}
\newcommand{\omC}[1]{\parbox[t]{0.09\linewidth}{\centering #1}}

\begin{table}[htbp]
\centering
\caption{Constraints on $\Omega_\mathrm{m}$ in three cosmological models. The column ``range'' denotes the best confidence level that contains the best-fit value from the combined (``ALL'') dataset.}
\label{tab:omega_m}
\setlength{\tabcolsep}{0.6mm}
\renewcommand{\arraystretch}{1.3}
\small
\begin{tabular}{l|l|llll|l}
\hline\hline
\omA{Model/Data} & \omB{best-fit} & \omC{$1\sigma$ lower} & \omC{$1\sigma$ upper} & \omC{$2\sigma$ lower} & \omC{$2\sigma$ upper} & \omC{level} \\
\hline
\omA{$\mathbf{NMCQ}$} & \omB{$0.305$} & \omC{} & \omC{} & \omC{} & \omC{} & \omC{} \\
\omA{\texttt{Planck 2018} CMB} & \omB{$0.31108$} & \omC{$0.279$} & \omC{$0.342$} & \omC{$0.243$} & \omC{$0.384$} & \omC{$<1\sigma$} \\
\omA{\texttt{DESI DR2} BAO} & \omB{$0.286$} & \omC{$0.275$} & \omC{$0.298$} & \omC{$0.262$} & \omC{$0.308$} & \omC{$<2\sigma$} \\
\omA{\texttt{DESY5} SN Ia} & \omB{$0.304$} & \omC{$0.251$} & \omC{$0.335$} & \omC{$0.140$} & \omC{$0.360$} & \omC{$<1\sigma$} \\
\hline
\omA{$\mathbf{w_0w_a CDM}$} & \omB{$0.319$} & \omC{} & \omC{} & \omC{} & \omC{} & \omC{} \\
\omA{\texttt{Planck 2018} CMB} & \omB{$0.150$} & \omC{$0.139$} & \omC{$0.354$} & \omC{$0.136$} & \omC{$0.757$} & \omC{$<1\sigma$} \\
\omA{\texttt{DESI DR2} BAO} & \omB{$0.390$} & \omC{$0.343$} & \omC{$0.438$} & \omC{$0.291$} & \omC{$0.487$} & \omC{$<2\sigma$} \\
\omA{\texttt{DESY5} SN Ia} & \omB{$0.504$} & \omC{$0.460$} & \omC{$0.537$} & \omC{$0.389$} & \omC{$0.572$} & \omC{$<3\sigma$} \\
\hline
\omA{$\mathbf{\Lambda CDM}$} & \omB{$0.302$} & \omC{} & \omC{} & \omC{} & \omC{} & \omC{} \\
\omA{\texttt{Planck 2018} CMB} & \omB{$0.315$} & \omC{$0.309$} & \omC{$0.322$} & \omC{$0.303$} & \omC{$0.328$} & \omC{$<3\sigma$} \\
\omA{\texttt{DESI DR2} BAO} & \omB{$0.298$} & \omC{$0.289$} & \omC{$0.306$} & \omC{$0.280$} & \omC{$0.316$} & \omC{$<1\sigma$} \\
\omA{\texttt{DESY5} SN Ia} & \omB{$0.352$} & \omC{$0.334$} & \omC{$0.368$} & \omC{$0.321$} & \omC{$0.386$} & \omC{$>3\sigma$} \\
\hline\hline
\end{tabular}
\end{table}

\section{Evidence for nonminimal coupling}

\begin{figure*}[t]
  \includegraphics[width=0.49\linewidth]{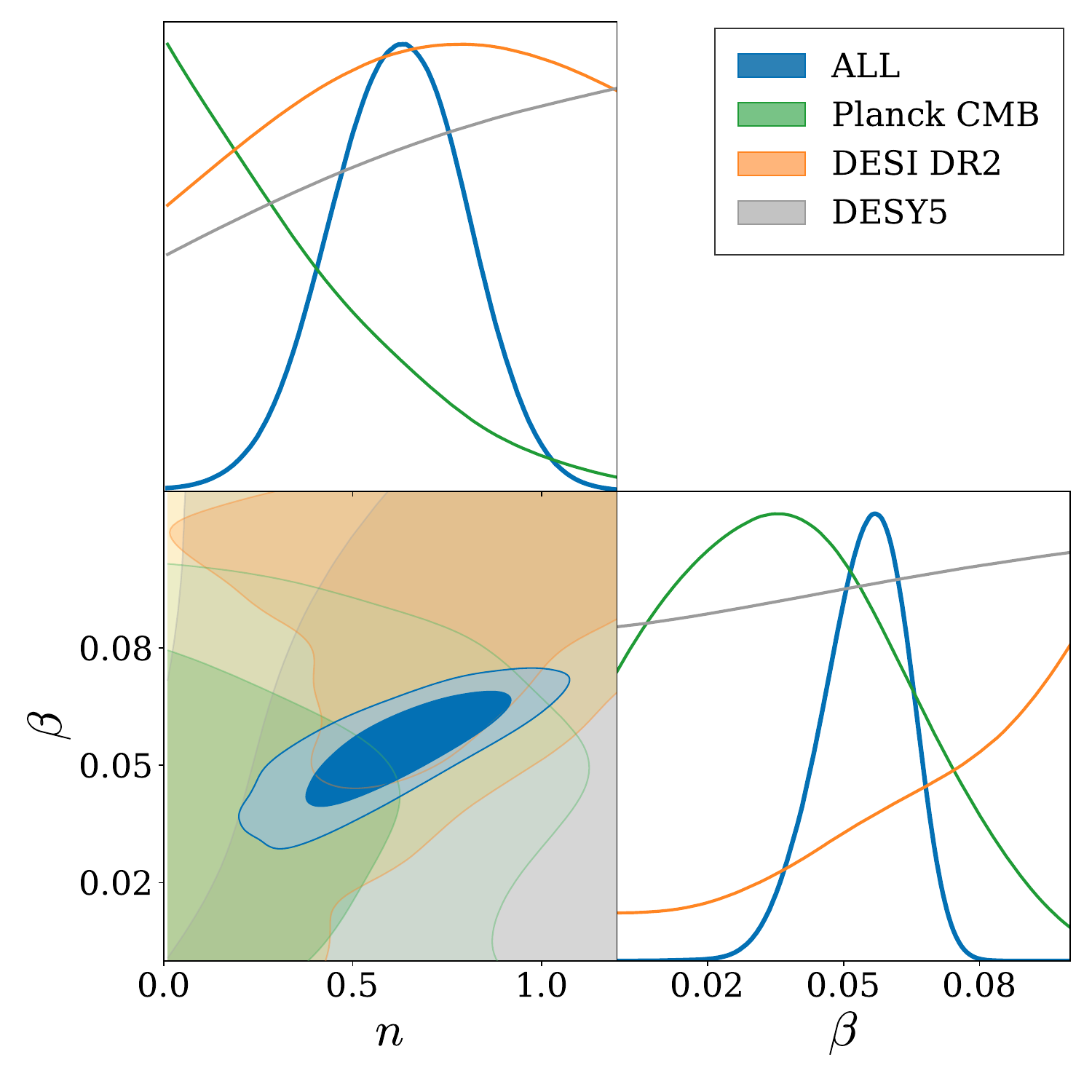}
  \includegraphics[width=0.49\linewidth]{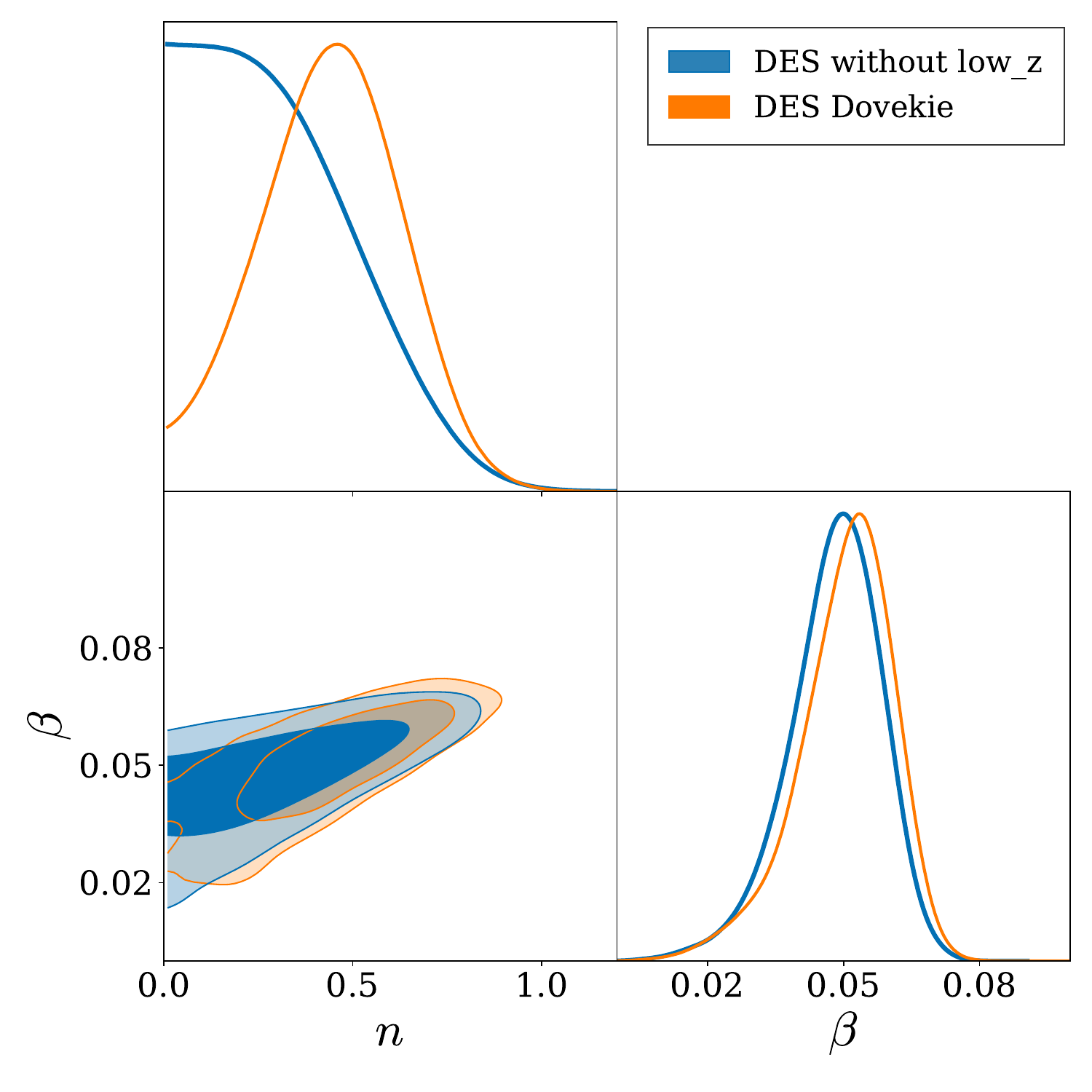}\\
  \caption{\label{fig:nbeta} Left: cosmological constraints on the model parameters $n$ and $\beta$ in the NMCQ model from \textit{Planck} 2018 CMB, DESI Y3 BAO DR2, DESY5 SNe Ia, and their combined datasets. Right: cosmological constraints on the model parameters $n$ and $\beta$ in the NMCQ model from Planck-CMB, DESI DR2 BAO, and DESY5 without low-$z$ samples (blue) or DES Dovekie (orange).}
\end{figure*}

The distribution of two model parameters, $n,\beta$ is shown in the left panel Fig.~\ref{fig:nbeta}. Notably, the evidence for the existence of nonzero $n,\beta$ is over $3\sigma$, and the constraint from each dataset overlaps in $1\sigma$. Further considering that the preference for $w_0w_a$CDM is noticeably reduced when the external low-$z$ SNe Ia in DESY5 data is discarded~\cite{Zhong:2025gyn}, we have re-run our code for the NMCQ model after removing the low-$z$ SN Ia samples in the combined CMB+BAO+SN analysis. As shown in the right panel of Fig.~\ref{fig:nbeta} below, the evidence for a nonzero $n$ becomes weaker after the low-$z$ SN data are discarded. This is expected since the late-time dynamical DE behavior is primarily constrained by the low-$z$ SN samples. On the other hand, the evidence for a nonzero $\beta$ remains stable at approximately $3\sigma$ level, essentially unchanged compared to our baseline analysis. Recently, the DES collaboration has improved their SN analysis in the released DES-Dovekie SN recalibration~\cite{DES:2025sig}, which has reduced the preference for $w_0w_a$CDM by $1\sigma$. Here, we have also updated the parameter constraint with this new dataset in the right panel of Fig.~\ref{fig:nbeta}, where the constraint on $\beta$ remains essentially unchanged, and the evidence for a nonzero $n$ is stronger than the above analysis without low-$z$ in DESY5, although slightly weaker (but still persists at roughly $2\sigma$ level) than our original constraint with full DESY5 data.

\bibliography{refer}

\end{document}